\documentclass[oneside,english]{elsarticle}
\usepackage[T1]{fontenc}
\usepackage{amsmath}
\usepackage{graphicx}
\usepackage{array}
\usepackage{longtable}

\providecommand{\\}{\\}

\providecommand{\openone}{\leavevmode\hbox{\small1\kern-3.5pt\normalsize1}}

\usepackage{babel}

\begin{document}
\begin{frontmatter}

\title{\hfill{\small{MPP-2009-142}}\\Chiral corrections to the $f_{V}^{{\perp}}/f_{V}$ ratio for vector mesons}

\author{Oscar Cat\`a}

\address{INFN, Laboratori Nazionali di Frascati, Via E. Fermi 40 I-00044 Frascati,
Italy}

\ead{ocata@lnf.infn.it}

\author{Vicent Mateu}

\address{Max-Planck-Institut f\"ur Physik (Werner-Heisenberg-Institut),\\
F\"ohringer Ring 6, 80805 M\"unchen, Germany}

\ead{mateu@mppmu.mpg.de}
\begin{abstract}
\noindent In this letter we compute the leading chiral corrections
to the ratio between the tensor and the vector decay couplings for the lowest
lying vector meson multiplet ($\rho$, $K^*$, $\phi$). We show that the leading chiral
logarithms arise from tadpole contributions and are therefore entirely fixed by chiral symmetry, while the next to leading corrections are purely analytic. Interestingly, the flavour structure of the chiral logarithms implies that only the $\rho$ meson is sensitive to pion logarithms. By examining $SU(2)_L\times SU(2)_R\times U(1)_S$ theories we show that this result can be easily understood and that it holds to all orders in the $m_s$ expansion. Recent lattice data seem to comply with our results.
\end{abstract}
\end{frontmatter}

\section{Introduction}

The coupling of the light vector mesons to the tensor current $f_V^{\perp}$ has
driven a lot of attention in the last years, the reason being
the crucial role that this coupling plays in the determination of
the $|V_{ub}|$ CKM parameter from exclusive semileptonic and radiative decays of
the $B$-meson within light-cone QCD sum rules (LCSR)~\cite{Ball:2006eu}. 

However, while the vector form factor can be extracted from experiments on
$e^{+}\, e^{-}$ annihilation or hadronic $\tau$ decays, the tensor
form factor can only be obtained theoretically. Since form factors are purely non-perturbative quantities, lattice gauge theory and effective field
theories are in this case the appropriate tools. 

On the effective field theory side, it is worth mentioning some attempts to estimate the value of $f_{\rho}^{{\perp}}$ in the chiral limit. Within
the context of Resonance Chiral Theory, in the strict large-$N_{C}$
limit and under the minimal hadronic ansatz approximation, Refs.~\cite{Mateu:2007tr}
and \cite{Jamin:2008rm} estimated at $\mathcal{O}(\alpha_{s}^{0})$
and $\mathcal{O}(\alpha_{s})$, respectively, the value of $f_{\rho}^{\perp}$.
Their result is compatible with the independent determination
of Ref.~\cite{Ball:2002ps}, in which LCSR were used. In Ref.~\cite{Cata:2008zc}
we derived a theorem valid at leading order in $1/N_{C}$ and $\alpha_{s}$
in which, for highly excited vector mesons, the ratio $f_{V}^{\perp}/f_{V}$
tends to $1/\sqrt{2}$ in absolute value. This value is surprisingly close to the lattice result for the
$\rho$ meson \cite{Becirevic:2003pn}, and seems to hint at some sort of universality for the ratio $f_V^{\perp}/f_V$.   

On the lattice side, it was already pointed out in Ref.~\cite{Becirevic:2003pn} that the light-cone sum rules in $B\rightarrow \rho\, \ell\, \nu$ are particularly sensitive to $f_{\rho}^{\perp}$ and therefore an effort from the lattice community to reduce its uncertainty was well motivated. Since then, many lattice collaborations have devoted studies to $f_V^{\perp}$ using different techniques and approximations (see Refs.~\cite{Becirevic:2003pn, Braun:2003jg, Gockeler:2005mh, Donnellan:2007xr, Dimopoulos:2008ee, Dimopoulos:2008hb, Allton:2008pn, Jansen:2009hr}), with overall agreement. 

One of the main sources of uncertainty in the previous lattice simulations is the extrapolation from the quark masses used in the simulations to the actual physical ones. The appropriate theoretical framework to perform such extrapolations is Chiral Perturbation Theory ($\chi$PT). Formulas for the decay couplings of pseudo-Goldstone bosons (using $\chi$PT) and B mesons (combining $\chi$PT and HQET) exist. For light vector mesons the situation is slightly more complicated\,: vector masses are not protected by chiral symmetry and their couplings to pseudo-Goldstone bosons spoil the power-counting of the theory. The power-counting can however be preserved if a heavy meson mass expansion is performed. This last approach, normally referred to as Heavy Meson Effective Theory (HMET) was introduced in Ref.~\cite{Jenkins:1995vb} to study the chiral corrections to the vector meson mass spectrum. The formalism was later extended to include external vector sources and used to evaluate the chiral corrections to $f_V$. In this letter we will generalize the theory to account for external tensor couplings in order to compute the chiral corrections to $f_V^{\perp}$.

In particular, we will compute the leading chiral logarithms directly to the ratio $f_V^{\perp}/f_V$. This will prove to be extremely convenient for several reasons\,: (i) all contributions to the wave function renormalization and mass corrections can be ignored, since they identically cancel in the ratio; (ii) we can make direct comparison with the lattice results, since this is precisely the quantity determined by the lattice groups; (iii) in lattice simulations, systematic uncertainties cancel in the ratio. In our effective theory approach, this translates into a reduction of the unknown low energy couplings that enter the calculation, thus enhancing the predictive power of our results.
 
This letter is organized as follows\,: in Section~\ref{Theoreticalframework} we discuss the introduction of external tensor sources in HMET and write down the Lagrangian relevant for the computation of the chiral corrections $f_V^{\perp}/f_V$ at leading order. In Section~\ref{Results} we present our results for two- and three-flavour HMET, while in Section~\ref{Newtheories} we discuss the case of $SU(2)_L\times SU(2)_R\times U(1)_S$ HMET, relevant when the strange quark is considered as heavy, by building effective field theories for the $K^*$ and the $\phi$. We finally present our conclusions in Section~\ref{conclusions}.

\section{Theoretical framework}
\label{Theoreticalframework}
In order to compute the chiral corrections to vector meson decay couplings we need an effective field theory describing the interactions of vector mesons with low-momentum ({\emph{i.e.}}, $p<\!\!< \Lambda_{\chi}\sim 1\,\,{\mathrm{GeV}}$) Goldstone bosons. A generic problem with such theories is that vector masses
do not vanish in the chiral limit and therefore derivatives on the vector fields are not soft and spoil the chiral power-counting. 

Since $m_V\sim \Lambda_{\chi}$, a possible solution is to split the vector momentum into a hard part and a residual one, $p_{\mu}=m \,v_{\mu}+k_{\mu}$, and perform a heavy mass expansion. Such an approach was introduced in Ref.~\cite{Jenkins:1995vb} under the name of HMET. The theory has been used in the past to compute corrections
to the vector meson masses~\cite{Jenkins:1995vb,Bijnens:1997ni}
and vector decay couplings~\cite{Bijnens:1998di}. Here we will extend the
formalism to include the couplings to tensor sources.

Vector fields in HMET are constructed from the fully relativitic ones by factoring out the hard momentum shells 
\begin{equation}
S_{\mu}\,=\,\dfrac{1}{\sqrt{2\, m}}\left[e^{-\,i\,m\,v\cdot x}\,{\hat{S}}_{\mu}^{(v)}+e^{\,i\,m\,v\cdot x}\,{\hat{S}}_{\mu}^{(v)\,\dagger}\right]+{\hat{S}}_{\mu}^{(v)\,||}\,,\label{eq:Heavy-limit}
\end{equation}
such that the effective fields ${\hat{S}}_{\mu}^{(v)}$, which satisfy $v\cdot {\hat{S}}_{\mu}^{(v)}=0$, only depend on the residual momentum $k$. In the previous equation, $m$ stands for the mass of the vector mesons in the chiral limit.  Since $k<\!\!< m \,v\sim \Lambda_{\chi}$, the chiral power-counting is clearly preserved. It is worth noting that the field ${\hat{S}}_{\mu}^{(v)}$
only contains the annihilation mode, and correspondingly ${\hat{S}}_{\mu}^{(v)\,\dagger}$
only the creation mode.{\footnote{${\hat{S}}_{\mu}^{||}$ can be shown to be a hard mode and therefore can be integrated out in the effective action.}} In the following we will drop the velocity label and refer to the vector fields simply as ${\hat{S}}_{\mu}$ and ${\hat{S}}_{\mu}^{\dagger}$.

In the framework of $SU(3)_L\times SU(3)_R$ $\chi$PT the vector fields ${\hat{S}}_{\mu}$ transform as
\mbox{${\hat{S}}_{\mu}\to h\, {\hat{S}}_{\mu}\, h^{\dagger}$}, such that under transformatios of the
unbroken $SU(3)_V$ group they behave as the adjoint representation. In our analysis we will assume ideal mixing,
such that the $\phi$ is a pure ${\bar{s}}s$ state, while the $\rho^0$ and the $\omega$ are orthogonal combinations 
of $u$ and $d$ quarks. Under this assumption the octet and singlet vector fields can be joined as a nonet
\begin{equation}
{\hat{S}}_{\mu}=\left[\begin{array}{ccc}
\frac{\omega+\rho^{0}}{\sqrt{2}} & \rho^{+} & K^{*+}\\[1.5ex]
\rho^{-} & \frac{\omega-\rho^{0}}{\sqrt{2}} & K^{*0}\\[3.5ex]
K^{*-} & \bar{K}^{*0} & \phi\end{array}\right]_{\mu}.
\end{equation}

\noindent As usual, Goldstone modes will be collected in a special unitary matrix $u$ transforming as \mbox{$u\to V_L\, u\, h^{\dagger}=h\, u\, V_R^{\dagger}$},
\begin{equation}
u\,=\,{\mathrm{exp}}\left(i\, \frac{\Pi}{\sqrt{2}\,f}\right)~;\qquad
\Pi\,=\,\left[\begin{array}{ccc}
{{\dfrac{\pi^0}{\sqrt{2}}}+{\dfrac{\eta}{\sqrt{6}}}} & \pi^+ & K^{+}\\[1.5ex]
\pi^- & {{ -\,\dfrac{\pi^0}{\sqrt{2}}}+{\dfrac{\eta}{\sqrt{6}}}} & K^{0}\\[3.5ex]
K^{-} & {\bar{K}}^{0} & -\,{2\,\dfrac{\eta}{\sqrt{6}}}
\end{array}\right]\,.
\end{equation}  

\noindent HMET can be extended to account for the interactions with arbitrary external sources. The theory was initially purported to describe interactions of vector mesons with soft pions and photons \cite{Jenkins:1995vb}. However, to describe a form factor we need an external source carrying enough
momenta to create a heavy resonance. In Ref.~\cite{Bijnens:1998di}
the theory was adapted to account for hard photons. The method can be easily extended to generic spin-one sources
(\emph{i.e.} vector, axial-vector, and tensor). 
  
Let $s$ be a generic spin-one external source, and let us decompose
it, by analogy to Eq.~(\ref{eq:Heavy-limit}), into hard and soft components
\begin{equation}\label{generic}
s\,=\,e^{-\,i\,m\,v\cdot x}\,\hat{s}+e^{i\,m\,v\cdot x}\,\hat{s}^{\dagger}+\tilde{s}\,,
\end{equation}
where $\hat{s}$ creates a vector meson, $\hat{s}^{\dagger}$ annihilates it, and $\tilde{s}$ is a soft interaction. One then has the conventional (soft) chiral transformations, but also hard transformations, with support over hard momentum shells. Chiral Ward identities are satisfied if under a chiral transformation $\tilde{\ell}$, $\tilde{r}$
transform locally and $\hat{\ell}$, $\hat{r}$ globally, whereas under a hard transformation, $\tilde{\ell}$, $\tilde{r}$ stay invariant and $\hat{\ell}$, $\hat{r}$ transform locally (see Ref.~\cite{Bijnens:1998di} for
details).

For tensor sources, a decomposition like Eq.~(\ref{generic}) also follows, but chiral invariance requires the presence of two fields $t_{LR}^{\mu\nu}$ and $t_{RL}^{\mu\nu}$, transforming as
\begin{equation}
\{\,{\hat{t}}_{LR}^{\mu\nu}\,,\,{\hat{t}}_{RL}^{\mu\nu}\,\}\,\,\longmapsto\,\, \{\,V_L\, {\hat{t}}_{LR}^{\mu\nu}\,V_R^{\dagger}\,,\,V_R\, {\hat{t}}_{RL}^{\mu\nu}\, V_L^{\dagger}\,\}\,,
\end{equation} 
which are chiral projections of the tensor source
\begin{equation}
t^{\mu\nu}\,=\,P_+^{\mu\nu\lambda\rho} \,{\hat{t}}_{LR\,\lambda\rho}+ P_-^{\mu\nu\lambda\rho}\, {\hat{t}}_{LR\,\lambda\rho}\,,
\end{equation}
where\,{\footnote{Tensor sources were introduced in $\chi$PT in Ref.~\cite{Cata:2007ns}, and we refer the reader there for all the details concerning how to decompose the tensor source
in irreducible chiral operators.}}
\begin{equation}
P_{\pm}^{\mu\nu\lambda\rho}\,=\, \frac{1}{4}(g^{\mu\lambda}\,g^{\nu\rho}- g^{\mu\rho}\,g^{\nu\lambda}\pm i\, \epsilon^{\mu\nu\lambda\rho})\,.      
\end{equation}
 
Since there are no Ward identities associated to the tensor current, under a hard transformation ${\tilde{t}}_{\mu\nu}$ stays invariant and ${\hat{t}}_{\mu\nu}$ transforms globally, while they both transform globally under a chiral transformation.  
 
In order to build the Lagrangian, it is convenient to work in the following basis\,:
\begin{eqnarray}\label{basis}
u_{\mu}&=&i\,u^{\dagger}D_{\mu}U u^{\dagger}=i\,[u^{\dagger}(\partial_{\mu}-i\,r_{\mu})u-u(\partial_{\mu}-i\,\ell_{\mu})u^{\dagger}]\,,\\
\chi_{\pm}&=&u^{\dagger} \chi\, u^{\dagger}\pm u\, \chi^{\dagger} u\,,\nonumber
\end{eqnarray}
together with the following external sources for vector and tensor sources\,:
\begin{eqnarray}\label{extsources}
{\hat{Q}}_{\pm}^{\mu} = u\,\hat{r}^{\mu}u^{\dagger}\pm u^{\dagger}\hat{\ell}^{\mu}u\,\,\,\,\,\,,&\qquad& \,\,{\hat{Q}}_{\pm}^{\mu\,+} = u\,\hat{r}^{\mu\,\dagger}u^{\dagger}\pm u^{\dagger}\hat{\ell}^{\mu\,\dagger}u\,\,\,\,\,\,\,\,\,\,,\nonumber\\
\hat{T}_{\pm}^{\mu\nu} = u^{\dagger}\hat{t}_{LR}^{\mu\nu}\, u^{\dagger}\pm u\,\hat{t}_{RL}^{\mu\nu}\, u\,,&\qquad& \hat{T}_{\pm}^{\mu\nu\,+}=u^{\dagger}\hat{t}_{LR}^{\mu\nu\,+}\, u^{\dagger}\pm u\,\hat{t}_{RL}^{\mu\nu\,+}\, u\,,\label{eq:hard-covariant}
\end{eqnarray}
such that all elements transform under the chiral group as $X\,\to h\, X\, h^{\dagger}$. For our purposes, $\chi$ will only play the role of a mass insertion operator, and therefore $\chi_+=2\,\chi=4\,B_0\,{\mathrm{diag}}(m_u,m_d,m_s)$.

Since we have explicitly split vector fields and external sources into positive and negative frequency modes,
in HMET the CPT theorem is not automatically satisfied, and we will require
our effective Lagrangian to be separately invariant under C, P and
T. In particular T invariance will be essential to ensure reality of the low energy parameters of the theory~\cite{Bijnens:1998di}.
In Table~\ref{tab:transformation} we display the transformation
properties of all our building blocks under discrete symmetries.
As usual, we promote
the derivatives to covariant derivatives $\nabla$ such that $\nabla_{\mu}X$
transforms in the same way as $X$ (the definitions can be found, \emph{e.g.}
in Ref~\cite{Cata:2008zc}\,).\renewcommand{\arraystretch}{1.4}\setlength{\LTcapwidth}{\textwidth}

\begin{table}[t]
\begin{center}
\begin{tabular}{|c|c|c|c|c|}
\hline 
 & $\mathcal{P}$  & $\mathcal{C}$  & $\mathcal{T}$  & h.c.\\
\hline 
${\hat{S}}^{\mu}$  & ${\hat{S}}_{\mu}$  & $-\, {\hat{S}}^{\mu\, T}$  & ${\hat{S}}_{\mu}$  & ${\hat{S}}^{\mu\,\dagger}$ \\
\hline 
$u^{\mu}$  & $-\, u_{\mu}$  & $u^{\mu\, T}$  & $u_{\mu}$  & $u^{\mu}$ \\
\hline 
$\chi_{\pm}$  & $\pm\,\chi_{\pm}$  & $\chi_{\pm}^{T}$  & $\chi_{\pm}$  & $\pm\,\chi_{\pm}$ \\
\hline 
$\hat{Q}_{\pm}^{\mu}$  & $\pm\,\hat{Q}_{\pm\,\mu}$  & $\mp\,\hat{Q}_{\pm}^{\mu}$  & $\hat{Q}_{\pm\,\mu}$  & $\hat{Q}_{\pm}^{\mu\,+}$\\
\hline 
$\hat{T}_{\pm}^{\mu\nu}$  & $\pm\,\hat{T}_{\pm\,\mu\nu}$  & $-\,\hat{T}_{\pm}^{\mu\nu}$  & $-\,\hat{T}_{\pm}^{\mu\nu}$  & $\pm\,\hat{T}_{\pm}^{\mu\nu\,+}$\\
\hline
$\nabla^{\mu}$ & $\nabla_{\mu}$ & $\nabla_{\mu}$ & $-\,\nabla_{\mu}$ & $\nabla^{\mu}$\\
\hline
$v^{\mu}$ & $v_{\mu}$ & $v^{\mu}$ & $v_{\mu}$ & $v^{\mu}$\\
\hline
\end{tabular}
\end{center}

\caption{Transformation properties of the HMET building blocks under discrete
symmetries.\label{tab:transformation}}

\end{table}
 
Before we write down the Lagrangian it is worth commenting on the chiral power-counting of the theory. The fact that HMET provides a consistent power-counting means that there is a well-defined relation between each diagram and its scaling with soft momentum and masses. This turns out to be a very efficient tool to list down all the diagrams contributing to a given order in the chiral expansion. The explicit formula can be cast as
\begin{equation}
D\,=\,2\,+\,2\, N_{L}\,+\, N_{R}\,+\,\sum_{n}\mathcal{N}_{n}(n-2)\,,\label{eq:Weinberg}
\end{equation}
where $D$ is the chiral order $p^D$ of the diagram, $N_{L}$ is the number of loops, $N_{R}$ is the number of resonance
internal lines and $\mathcal{N}_{n}$ is the number of vertices coming
from the $\mathcal{L}_{n}$ Lagrangian. Eq.~(\ref{eq:Weinberg}) is the analog in HMET of the familiar Weinberg's power-counting formula in $\chi$PT. Indeed, setting $N_R=0$ one recovers Weinberg's formula, except for the fact that in general in HMET we have operators with odd chiral counting.  

In full generality, the contributions to the decay couplings $f_V$ and $f_V^{\perp}$ at the quantum level can be classified in\,: (i) mass corrections; (ii) wave-function renormalization and (iii) vertex renormalization.  The first two contributions are multiplicative and identically cancel in the ratios $f_V^{\perp}/f_V$ to all orders in the chiral expansion. Their explicit expressions can be found in Refs.~\cite{Jenkins:1995vb, Bijnens:1997ni}. Vertex renormalization, on the contrary, will account for the fact that the vector and tensor sources, ${\hat{\ell}}^{\mu}$ ${\hat{r}}^{\mu}$,  and ${\hat{t}}_{LR}^{\mu\nu}$, ${\hat{t}}_{RL}^{\mu\nu}$ transform differently under the chiral group.

For the sake of simplicity, we will assign no chiral counting to the hard external sources and the vector fields. Then the leading Lagrangian relevant for our computation reads  
\begin{eqnarray}\label{Laglead}
{\cal{L}}_{0}&=& -\,i\, \langle {\hat{S}}_{\mu}^{\dagger}(v\cdot\nabla) {\hat{S}}^{\mu}\rangle+ \lambda_1 \langle {\hat{S}}_{\mu} {\hat{Q}}_{+}^{\mu\,\dagger}\rangle +\,i\, \lambda_2 \langle {\hat{S}}^{\mu} \,{\hat{T}}_{+\mu\nu}^{\dagger}\rangle v^{\nu}\nonumber\\
&&+\;\lambda_3 \langle {\hat{S}}_{\mu}\rangle\,\langle {\hat{Q}}_+^{\mu\, \dagger}\rangle +\,i\, \lambda_4\, \langle {\hat{S}}^{\mu}\rangle\,\langle {\hat{T}}_{+\mu\nu}^{\dagger}\rangle\, v^{\nu}+{\mathrm{h.c.}}\,,
\end{eqnarray}
Ideal mixing, which we assumed when defining our vector fields, turns out to be a good phenomenological approximation, but it also allows us to endow the theory with a large-$N_c$ power-counting. For our purposes, this will only entail that operators with multiple traces in the Lagrangian, which are Zweig-suppressed, will be also $1/N_c$-suppressed. In particular, such $1/N_c$ effects in Eq.~(\ref{Laglead}) correspond
to the singlet component of the external sources\,\footnote{We take nonets of currents\,: $\langle {\hat{Q}}_{\pm}^{\mu}\rangle \neq0\neq \langle {\hat{T}}_{\pm}^{\mu\nu}\rangle$.}
and therefore will only affect the decay couplings of the $\omega$ and $\phi$. 

The next to leading order operators are
\begin{eqnarray}
{\cal{L}}_{1}&=&\quad\epsilon^{\mu\nu\rho\lambda}\left[\,i\,g_1\,\langle u_{\mu}\{ {\hat{S}}_{\rho},{\hat{S}}_{\lambda}^{\dagger}\}\, \rangle\, v_{\nu}
+\,i\, g_2\,\langle u_{\lambda}\{ {\hat{S}}_{\mu},{\hat{Q}}_{+\,\rho}^{\dagger}\} \rangle\, v_{\nu}\right.\nonumber\\
&&+\left. g_3\,\langle u_{\mu}\{ {\hat{S}}_{\rho},{\hat{T}}_{+\nu\lambda}^{\dagger}\} \rangle
+\,i\,g_4\,\{\langle u_{\mu}\, {\hat{S}}_{\rho}\rangle \langle{\hat{S}}_{\lambda}^{\dagger}\rangle+\langle u_{\mu}\, {\hat{S}}_{\lambda}^{\dagger}\rangle \langle{\hat{S}}_{\rho}\rangle\}\, v_{\nu}\right.\nonumber\\
&&+\left.\,i\, g_5\,\langle u_{\lambda}\, {\hat{S}}_{\mu}\rangle\,\langle {\hat{Q}}_{+\,\rho}^{\dagger} \rangle\, v_{\nu}+\,i\, g_6\, \langle u_{\lambda}\, {\hat{Q}}_{+\,\rho}^{\dagger}\rangle\,\langle {\hat{S}}_{\mu}\rangle\, v_{\nu}\right.\nonumber\\
&&+\left.g_7\,\langle u_{\mu}\, {\hat{S}}_{\rho}\rangle\,\langle {\hat{T}}_{+\nu\lambda}^{\dagger} \rangle+g_8\,\langle u_{\mu}\,{\hat{T}}_{+\nu\lambda}^{\dagger}\rangle\,\langle {\hat{S}}_{\rho} \rangle \right]+{\mathrm{h.c.}}\,,\label{LagNLO}
\end{eqnarray}
and
\begin{eqnarray}
{\cal{L}}_{2}&=& \quad\!\mu_1 \,\langle \{ {\hat{S}}^{\mu},\chi_+\}\,{\hat{Q}}_{+\,\mu}^{\dagger}\rangle +\,i\, \mu_2\, \langle \{ {\hat{S}}^{\mu},\chi_+\}\,{\hat{T}}_{+\mu\nu}^{\dagger}\rangle \,v^{\nu}\nonumber\\
&&+\,\mu_3\, \langle {\hat{S}}^{\mu}\, \chi_+\rangle\,\langle {\hat{Q}}_{+\,\mu}^{\dagger}\rangle +\,i\, \mu_4\, \langle {\hat{S}}^{\mu}\, \chi_+\rangle\,\langle {\hat{T}}_{+\mu\nu}^{\dagger}\rangle\, v^{\nu}\nonumber\\
&&+\,\mu_5 \,\langle {\hat{S}}^{\mu}\rangle\,\langle \chi_+\, {\hat{Q}}_{+\,\mu}^{\dagger}\rangle +\,i\, \mu_{6} \,\langle {\hat{S}}^{\mu}\rangle\,\langle \chi_+\, {\hat{T}}_{+\mu\nu}^{\dagger}\rangle\, v^{\nu}\nonumber\\
&&+\,\mu_{7}\, \langle \chi_+\rangle\,\langle {\hat{S}}^{\mu}\, {\hat{Q}}_{+\,\mu}^{\dagger}\rangle +\,i\, \mu_{8}\, \langle \chi_+\rangle\, \langle {\hat{S}}^{\mu}\, {\hat{T}}_{+\mu\nu}^{\dagger}\rangle\, v^{\nu}\nonumber\\
&&+\,\mu_{9} \,\langle {\hat{S}}^{\mu}\rangle\,\langle \chi_+\rangle\,\langle {\hat{Q}}_{+\,\mu}^{\dagger}\rangle +\,i\, \mu_{10}\, \langle {\hat{S}}^{\mu}\rangle\,\langle \chi_+\rangle\,\langle {\hat{T}}_{+\mu\nu}^{\dagger}\rangle\, v^{\nu}+{\mathrm{h.c.}}\,,\label{LagNNLO}
\end{eqnarray}
The quantities we want to compute are defined as 
\begin{eqnarray}
\langle\,0\,|\,\bar{u}\,\gamma_{\mu}\, d(0)\,|\,\rho^{+}(p,\lambda)\,\rangle & = & f_{\rho}\, m_{\rho}\,\epsilon_{\mu}^{(\lambda)}\,,\label{eq:vector-definition}\nonumber\\
\langle\,0\,|\,\bar{u}\,\sigma_{\mu\nu}\, d(0)\,|\,\rho^{+}(p,\lambda)\,\rangle &=& i\, f_{\rho}^{\perp}\, m_{\rho}\,(\epsilon_{\mu}^{(\lambda)}\, v_{\nu}-\epsilon_{\nu}^{(\lambda)}\, v_{\mu})\,,
\end{eqnarray}
where we should note that the couplings are defined to be related to isospin currents. Therefore, it follows that
\begin{eqnarray}\label{rem}
\langle\,0\,|\,\bar{u}\,\gamma_{\mu}\, s(0)\,|\, K^{*+}(p,\lambda)\,\rangle & = & f_{K}\, m_{K}\,\epsilon_{\mu}^{(\lambda)}\,,\nonumber \\
\langle\,0\,|\,\bar{s}\,\gamma_{\mu}\, s(0)\,|\,\phi(p,\lambda)\,\rangle & = & f_{\phi}\, m_{\phi}\,\epsilon_{\mu}^{(\lambda)}\,,\nonumber \\
\dfrac{1}{\sqrt{2}}\,\langle\,0\,|\left[\bar{u}\,\gamma_{\mu}\, u(0)+\bar{d}\,\gamma_{\mu}\, d(0)\right]|\,\omega(p,\lambda)\,\rangle & = & f_{\omega}\, m_{\omega}\,\epsilon_{\mu}^{(\lambda)}\,. 
\end{eqnarray}
The definitions for the remaining members of the nonet can be easily inferred from Eqs.~(\ref{eq:vector-definition}) and (\ref{rem}) above.

In terms of the effective field theory, formally we need to compute the following matching equation\,:
\begin{equation}
\langle \,0\,|\,\frac{i\,\delta S_{QCD}}{\delta J_{\rho}}\,|\, \rho^{+}(p,\lambda)\,\rangle =\langle\,0\,|\,\frac{i\,\delta S_{eff}}{\delta J_{\rho}}\,|\, \rho^{+}(p,\lambda)\,\rangle \equiv f_{\rho} m_{\rho} \epsilon_{\mu}\,,
\end{equation}
(and a similar one for $f_{\rho}^{\perp}$)
for each decay coupling, where $J_{\rho}$ is the external vector current with the flavour content to excite a $\rho$ field.

In the following we will work in the isospin limit, setting $m_u=m_d=\overline{m}$. Accordingly, we will pick the states in the previous equations as representatives for each isospin multiplet. In this limit, together with the assumption of ideal mixing, mixing between neutral states can be ignored altogether.
 
Moreover, the Goldstone boson mass matrix $\chi_{+}$, using the Gell-Mann--Okubo formula, can be cast as
\begin{equation}
\chi_{+}=2\left(\begin{array}{ccc}
m_{\pi}^{2} & 0 & 0\\
0 & m_{\pi}^{2} & 0\\
0 & 0 & 2\, m_{K}^{2}-m_{\pi}^{2}\end{array}\right)\,.
\end{equation} 

\section{Results}
\label{Results}
\begin{figure}[t]
\begin{center}
\includegraphics[width=10cm]{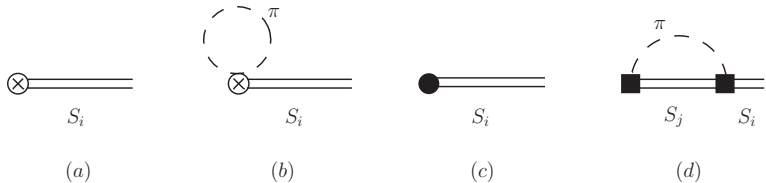}
\end{center}
\caption{Diagrams contributing to the ratio $f_V^{\perp}/f_V$ up to $m_q^{3/2}$ in HMET. From left to right, LO [(a)], NLO [(b) and (c)] and NNLO [(d)] contributions. Circle-cross  vertices depict operators of ${\cal{O}}(1)$, while box vertices and dotted vertices represent ${\cal{O}}(p)$ and ${\cal{O}}(p^2)$ operators, respectively.  \label{fig:Diagrams} }
\end{figure}

Using Eq.~(\ref{eq:Weinberg}) the relevant diagrams are displayed in Fig.~\ref{fig:Diagrams}, where we have only included the non-vanishing contributions to $f_V^{\perp}/f_V$, {\emph{i.e.}}, the vertex renormalization contributions. The tadpole in Fig.~\ref{fig:Diagrams}(b) is the leading chiral correction, at $\mathcal{O}(p^{2})$, and it is determined entirely by the leading order operators of Eq.~(\ref{Laglead}). The resulting non-analytic mass dependence is entirely fixed by chiral symmetry, and amounts to a renormalization of the parameters of Eq.~(\ref{Laglead}). Fig.~\ref{fig:Diagrams}(c) comes from operators in Eq.~(\ref{LagNNLO}) and is requested to absorb the divergences of Fig.~\ref{fig:Diagrams}(b) and render the final result finite. The sunset diagram of Fig.~\ref{fig:Diagrams}(d), on the other hand, is the $\mathcal{O}(p^{3})$ contribution coming from the operators in Eq.~(\ref{LagNLO}), and it can be shown to be finite~\cite{Jenkins:1995vb,Bijnens:1998di}.

Higher order operators will generate tadpole and sunset diagrams that will contribute at higher orders in the chiral expansion. However, it can be easily shown that there is no ${\cal{O}}(p^3)$ tadpole. Therefore, the ${\cal{O}}(p^3)$ corrections to each separate decay coupling will depend on a sizeable number of low energy parameters, but will yield an analytic contribution.

In the following we will present the results at ${\cal{O}}(p^2)$, which contain the non-analytic terms coming from Fig.~\ref{fig:Diagrams}(b) together with the corresponding counterterms in Fig.~\ref{fig:Diagrams}(c), 
for the case of two and three dynamical flavours.

For $SU(2)$, a number of simplifications take place. On the one hand, some of the parameters in the Lagrangian are linearly related. Using the Cayley-Hamilton relations, one can show that the following sets of parameters\,: $\{g_1,g_4\}$, $\{g_3,g_7,g_8\}$, $\{g_2,g_5,g_6\}$, $\{\mu_1,\mu_3,\mu_5,\mu_7,\mu_9\}$ and $\{\mu_2,\mu_4,\mu_6,\mu_8,\mu_{10}\}$ are related and therefore one coupling in each subset can be eliminated. Moreover, in the isospin limit $\chi_+=2\,m_{\pi}^2\, \openone_2$, which means that only one representative of the subsets $\{\mu_3,\mu_5,\mu_9\}$ and $\{\mu_4,\mu_6,\mu_{10}\}$ is independent.

Taking all these considerations into account, the results for the $SU(2)$ analysis are the following\,:
\begin{eqnarray}\label{su2}
\frac{f_{\rho}^{\perp}}{f_{\rho}}&=& \frac{f_V^{\perp}}{f_V}\left[1+2\,m_{\pi}^2\,\lambda(\mu)+\frac{m_{\pi}^2}{32\,\pi^2 \,f^2}\log{\left(\frac{m_{\pi}^2}{\mu^2}\right)}\right]\nonumber\\
&\equiv&\frac{f_V^{\perp}}{f_V}\left[1+\frac{m_{\pi}^2}{32\,\pi^2\, f^2}\log{\left(\frac{m_{\pi}^2}{{\hat{m}}_{\rho}^2}\right)}\right]\,,\nonumber\\
\frac{f_{\omega}^{\perp}}{f_{\omega}}&=&\frac{f_V^{\perp}}{f_V}\left[1+2\,\gamma_C+2\,m_{\pi}^2\,[{\bar{\lambda}}(\mu)+\lambda(\mu)]-\frac{3\,m_{\pi}^2}{32\,\pi^2\,f^2}\log{\left(\frac{m_{\pi}^2}{\mu^2}\right)}\right]\nonumber\\
&\equiv&\frac{f_V^{\perp}}{f_V}\left[1+2\,\gamma_C-\frac{3\,m_{\pi}^2}{32\,\pi^2\, f^2}\log{\left(\frac{m_{\pi}^2}{{\hat{m}}_{\omega}^2}\right)}\right]\,,
\end{eqnarray}
where $f_V^{\perp}/f_V=\lambda_2/4\lambda_1$ and $\lambda$, ${\bar{\lambda}}$ (and correspondingly ${\hat{m}}_{\rho}$, ${\hat{m}}_{\omega}$) are combinations of couplings from ${\cal{L}}_{2}$. $\gamma_C$ is a Zweig-suppressed term coming from the second line of ${\cal{L}}_{0}$ and is expected to be extremely small.

For the $SU(3)$ case the results are
\begin{eqnarray}\label{results3}
\frac{f_{\rho}^{\perp}}{f_{\rho}}&=&\frac{F_V^{\perp}}{F_V}\left[1+(2\,\Lambda+{\bar{\Lambda}})\,m_{\pi}^2+2\,{\bar{\Lambda}}\,m_K^2+\frac{m_{\pi}^2}{32\,\pi^2\,f^2}\log{\left(\frac{m_{\pi}^2}{\mu^2}\right)}\right.\nonumber \\
&&\left. -\,\frac{m_{\eta}^2}{96\,\pi^2\,f^2}\log{\left(\frac{m_{\eta}^2}{\mu^2}\right)}\right]\,,\nonumber\\
\frac{f_K^{\perp}}{f_K}&=&\frac{F_V^{\perp}}{F_V}\left[1+{\bar{\Lambda}}\,m_{\pi}^2+2\,(\Lambda+{\bar{\Lambda}})\,m_K^2+\frac{m_{\eta}^2}{48\,\pi^2\,f^2}\log{\left(\frac{m_{\eta}^2}{\mu^2}\right)}\right]\,,\nonumber\\
\frac{f_{\omega}^{\perp}}{f_{\omega}}&=&\frac{F_V^{\perp}}{F_V}\left[1+2\,\Gamma_C+(2\,\Lambda+{\bar{\Lambda}}+2\,\Lambda_1)\,m_{\pi}^2+2\,{\bar{\Lambda}}\,m_K^2\right.\nonumber\\
&&\left.-\,\frac{3\,m_{\pi}^2}{32\,\pi^2\, f^2}\log{\left(\frac{m_{\pi}^2}{\mu^2}\right)}-\frac{m_{\eta}^2}{96\,\pi^2\,f^2}\log{\left(\frac{m_{\eta}^2}{\mu^2}\right)}\right]~\nonumber\\
\frac{f_{\phi}^{\perp}}{f_{\phi}}&=&\frac{F_V^{\perp}}{F_V}\left[1+\Gamma_C+({\bar{\Lambda}}-2\,\Lambda-\Lambda_1)\,m_{\pi}^2+2\,(2\,\Lambda+{\bar{\Lambda}}+\Lambda_1)\,m_K^2\right.\nonumber\\
&&\left.-\,\frac{m_{\eta}^2}{24\,\pi^2\,f^2}\log{\left(\frac{m_{\eta}^2}{\mu^2}\right)}\right]\,.
\end{eqnarray}
where $\Lambda$, ${\bar{\Lambda}}$, and $\Lambda_1$ are counterterms involving parameters from ${\cal{L}}_2$ which, just like in Eq.~(\ref{su2}), have a scale-dependence that absorbs the logarithmic divergence coming from the tadpole and renders the result finite. Again, just like in Eq.~(\ref{su2}), we used the $1/N_{C}$ expansion above for the $\phi$ and $\omega$, and have discarded terms of order $\mathcal{O}(m_{q}N_{C}^{-2})$. 

It is easy to check that the $SU(3)$ results reduce to the $SU(2)$ case when the strange mass is non-dynamical. Moreover, at the order we are working, one finds the following relation between the $SU(2)$ and $SU(3)$ couplings\,:
\begin{eqnarray}
\dfrac{f_{V}^{\perp}}{f_{V}}&=&\dfrac{F_{V}^{\perp}}{F_{V}}\left[1+2\,B_0\,m_s\,\bar{\Lambda}-
\dfrac{B_0\,m_s}{72\,\pi^{2}\,f^{2}}\,\log\left(\dfrac{4\,B_0\,m_s}{3\,\mu^{2}}\right)\right],\nonumber\\
\lambda &=& \Lambda+\bar{\Lambda}-\dfrac{1}{576\,\pi^{2}\,f^{2}}
\left[1+\log\left(\dfrac{4\,B_0\,m_s}{3\,\mu^2} \right)  \right],\nonumber\\
\bar{\lambda}&=&\Lambda_1\,,\qquad\gamma_C=\Gamma_C\,.
\end{eqnarray}

The flavour structure of the non-analytic terms in Eq.~(\ref{results3}) can be understood by looking at the structure of the tadpole vertex, which involves the operators $\langle {\hat{S}}_{\mu}\, {\hat{Q}}_{+}^{\mu\,\dagger}\rangle$ and $\langle {\hat{S}}^{\mu} \,{\hat{T}}_{+\mu\nu}^{\dagger}\rangle\, v^{\nu}$ with two Goldstone fields pulled out from ${\hat{Q}}_{+}^{\mu\,\dagger}$ and ${\hat{T}}_{+\mu\nu}^{\dagger}$. The only surviving contribution to $f_V^{\perp}/f_V$ is of the generic form ${\mathrm{Tr}}[{\hat{S}}\, \Pi\, {\cal{O}}\, \Pi]$, where ${\cal{O}}$ stands for either vector or tensor sources. It is easy to check that the contributions coming from such operator follow a block-diagonal structure\,: the upper $(2\times 2)$ $\rho-\omega$ sector in the vector matrix ${\hat{S}}$ will only receive contributions from the $(2\times 2)$ $\pi-\eta$ sector in the Goldstone matrix $\Pi$; $\phi$ only couples to $\eta$, because it is the only field in the $\Pi_{33}$ entry; while the $K^*$ has a non-vanishing contribution only because the $\eta$ field is present all along the diagonal in $\Pi$. In a similar fashion, one can show that the absence of kaon loops in any flavour channel is linked to the assumption of ideal mixing. Therefore, such an absence of kaon logarithms is not related to chiral symmetry and can be considered, to a certain extent, accidental.

It is worth mentioning that only $f_{\rho}^{\perp}/f_{\rho}$ and $f_{\omega}^{\perp}/f_{\omega}$ depend on pion logarithms [\,although each decay coupling in Eq.~(\ref{results3}) separately does\,]. This is especially interesting for the lattice\,: in many simulations only $m_u$ and $m_d$ are dynamical while $m_s$ is held fixed. If the structure of Eq.~(\ref{results3}) would hold even when one does not assume that the strange quark mass is small, this would entail that chiral logarithms only affect the $\rho(770)$, while the $K^*$ and $\phi$ should scale linearly with light quark masses.  

Eq.~(\ref{results3}), however, only shows the leading term in the strange quark mass expansion. Therefore, it is by construction insensitive to higher order powers of the strange quark mass that could potentially be associated with pion logarithms. In other words, with $SU(3)_L\times SU(3)_R$ HMET we cannot rule out contributions that scale like $m_{K,\eta}^n m_{\pi}^2\log{m_{\pi}^2}$. In order to do so we have to consider $SU(2)_L\times SU(2)_R\times U(1)_S$ HMET.

\section{A digression on $SU(2)_L\times SU(2)_R\times U(1)_S$ HMET}
\label{Newtheories}
In this section we will describe a family of effective field theories to study the interactions of the vector mesons with light up and down quarks when the strange quark is treated as heavy. Therefore, one has to study the interactions of the $\rho$, $K^*$ and $\phi$ with $SU(2)_L\times SU(2)_R$ Goldstone bosons in the presence of external sources. 

The first thing to notice is that vector mesons with different number of strange quarks decouple, {\emph{i.e.}}, there are no strangeness-changing interactions that can be mediated by the Goldstone bosons. Therefore, one can study separate effective theories for $SU(2)$ chiral triplets ($\rho$), doublets ($K^*$) and singlets ($\phi$). In the following we will obviate the strangeness quantum number, since the only relevant dynamical group is the chiral symmetry group. 

The effective theory for the $\rho$ meson is a chiral $SU(2)_L\times SU(2)_R$ theory, with sources
\begin{equation}
{\cal{L}}_{ext}={\cal{L}}_{QCD}+{\bar{q}}_R\gamma_{\mu}\,r^{\mu}q_R+
{\bar{q}}_L\gamma_{\mu}\,\ell^{\mu}q_L +
{\bar{q}}\sigma_{\mu\nu}t^{\mu\nu}q\,,\qquad {\bar{q}}=({\bar{u}},{\bar{d}})\,,
\end{equation}
in the adjoint representation and with $\rho_{\mu}\rightarrow h\,\rho_{\mu}\,h^{\dagger}$, {\emph{i.e.}}, conventional $SU(2)_L\times SU(2)_R$ HMET, with the building blocks of Eq.~(\ref{extsources}) and the results given in Eq.~(\ref{su2}).

For the $K^*$ one has to consider the following external sources
\begin{equation}
{\cal{L}}_{ext}={\cal{L}}_{QCD}+{\bar{s}}\gamma_{\mu}\,r^{\mu}q_R
+{\bar{s}}\gamma_{\mu}\,\ell^{\mu}q_L
+ {\bar{s}}\sigma_{\mu\nu}t^{\mu\nu}q+{\mathrm{h.c.}}\,,
\end{equation}
which transform in the fundamental representation of $SU(2)_L\times SU(2)_R$\,:\,\footnote{For the sake of simplicity, we have used the same notation for the external sources as in Section~\ref{Theoreticalframework}. However, it should be clear that they refer to different entities, as their transformations under the chiral group manifestly show.}  
\begin{eqnarray}
r_{\mu}&\rightarrow& r_{\mu} V_R^{\dagger}+i\,\partial_{\mu}V_R^{\dagger}\,,\nonumber\\
{\ell}_{\mu}&\rightarrow& {\ell}_{\mu} V_L^{\dagger}+i\,\partial_{\mu}V_L^{\dagger}\,,\nonumber\\
t^{\mu\nu}_R &\rightarrow& t^{\mu\nu}_R V_R^{\dagger}\,,\nonumber\\
t^{\mu\nu}_L &\rightarrow& t^{\mu\nu}_L V_L^{\dagger}\,.
\end{eqnarray}
Accordingly, the building blocks are
\begin{eqnarray}\label{build}
{\hat{Q}}_{\pm}^{\mu} = \hat{r}^{\mu}u^{\dagger}\pm \hat{\ell}^{\mu}u\,\,\,\,\,\,,&\qquad& \,{\hat{Q}}_{\pm}^{\mu\,+} = u\,\hat{r}^{\mu\,\dagger}\pm u^{\dagger}\hat{\ell}^{\mu\,\dagger}\,\,\,\,\,\,\,\,\,,\label{eq:hard-covariant-2}\\
\hat{T}_{\pm}^{\mu\nu} = \hat{t}_{R}^{\mu\nu}\, u^{\dagger}\pm \hat{t}_{L}^{\mu\nu}\, u\,,&\qquad& \hat{T}_{\pm}^{\mu\nu\,+}=u\,\hat{t}_{R}^{\mu\nu\,+}\pm u^{\dagger}\,\hat{t}_{L}^{\mu\nu\,+}\,,\nonumber 
\end{eqnarray}
which transform as $({\hat{Q}}^{\mu}_{\pm},{\hat{T}}^{\mu\nu}_{\pm})\rightarrow ({\hat{Q}}^{\mu}_{\pm},{\hat{T}}^{\mu\nu}_{\pm}) h^{\dagger}$. Correspondingly, the $K^*$ fields can be grouped in doublets 
\begin{equation}
K^*_{\mu}=\left(\begin{array}{c}
K^{+0}\\
K^{*0} 
\end{array}\right)_{\mu}\,,\qquad {\bar{K}}^*_{\mu}=({\bar{K}}^{-0},{\bar{K}}^{*0})_{\mu}\,,
\end{equation} 
transforming as $K^*_{\mu}\rightarrow h \,K_{\mu}^*$. 

For the $\phi$ field the external sources needed are
\begin{equation}
{\cal{L}}_{ext}={\cal{L}}_{QCD}+{\bar{s}}\gamma_{\mu}v_s^{\mu}s+ {\bar{s}}\gamma_{\mu}\,\gamma_5\,a_s^{\mu}s+ {\bar{s}}\sigma_{\mu\nu}t_s^{\mu\nu}s\,,
\end{equation}
which clearly are singlets under the chiral group. 

One could now build the Lagrangian for the $K^*$ and $\phi$ fields, in a similar fashion as what we already did in Section~\ref{Theoreticalframework} for the $SU(3)_L\times SU(3)_R$ HMET. However, for our discussion the important point to notice is that, contrary to the $\rho$ meson case, for the $K^*$ and $\phi$ the tensor and vector sources transform in the same way under the chiral group [\,as in Eq.~(\ref{build}) for the $K^*$ and trivially for $\phi$\,]. Therefore, chiral symmetry prevents the appearance of pion logarithms in the $K^*$ and $\phi$ channels\,: the contribution from pion loops, which is present in each decay coupling, will cancel once the ratio is taken. We emphasize that this result holds to all orders in the strange mass expansion, and in particular explains the absence of pion loops in the $K^*$ and $\phi$ channels in Eq.~(\ref{results3}).   
   
\section{Discussion and conclusions}
\label{conclusions}

In this paper we have computed the chiral corrections to the ratio $f_V^{\perp}/f_V$ for the lightest vector meson multiplet. We have performed our analysis using HMET, such that the $1/m_V$ expansion of the theory induces a well-defined chiral power counting. In order to compute $f_V^{\perp}/f_V$, we have introduced external tensor sources into the theory, along the lines described in Ref.~\cite{Cata:2007ns}, and built the relevant
terms of the Lagrangian which give rise to a non-trivial contribution to the ratio.

We find that the leading order correction is entirely determined by tadpole diagrams, and therefore is completely determined by chiral symmetry. The next to leading terms are ${\mathcal{O}}(m_{\pi}^3)$ and can be shown to be purely analytic. Eqs.~(\ref{su2}) and (\ref{results3}) are our final expressions to be fitted to lattice data. 
 
Taking the ratio of the decay couplings greatly simplifies the result\,: contributions from mass corrections and wave-function renormalization identically cancel. Moreover, kaon logarithms also cancel and pion logarithms only survive in the $\rho$ (and $\omega$) channels.

This last point is especially interesting for lattice simulations, since there one typically considers dynamical up and down quarks whereas the strange quark is treated as heavy. However, the results of Eq.~(\ref{results3}) are only valid at leading order in a (light) strange quark mass expansion, {\emph{i.e.}}, in $SU(3)_L\times SU(3)_R$ HMET, whereas a heavy strange quark requires a different framework, namely $SU(2)_L\times SU(2)_R\times U(1)_S$ HMET. In Section~\ref{Newtheories} we examined these set of effective field theories and concluded that there is absence of pion logarithms to all orders in the strange quark mass
. The key observation is that, contrary to $SU(3)_L\times SU(3)_R$ HMET, the tensor and vector sources transform in the same way, and therefore their interactions with pions cancel identically in the ratio. Therefore, $SU(2)_L\times SU(2)_R\times U(1)_S$ HMET predicts that only the $\rho$ should display the bending of the chiral logarithm, while the $K^*$ and $\phi$ should scale mainly like $m_{\pi}^2$, with subleading corrections of ${\cal{O}}(m_{\pi}^3)$ (and not $m_{\pi}^4$ as it is customarily assumed in lattice analyses). In other words, that the results for $f_V^{\perp}/f_V$ when the strange quark mass is heavy can be inferred from Eq.~(\ref{results3}) by naively freezing $m_s$. This is one of the main results of our paper. 

\begin{figure}[t]
\begin{center}
\includegraphics[width=9.0cm]{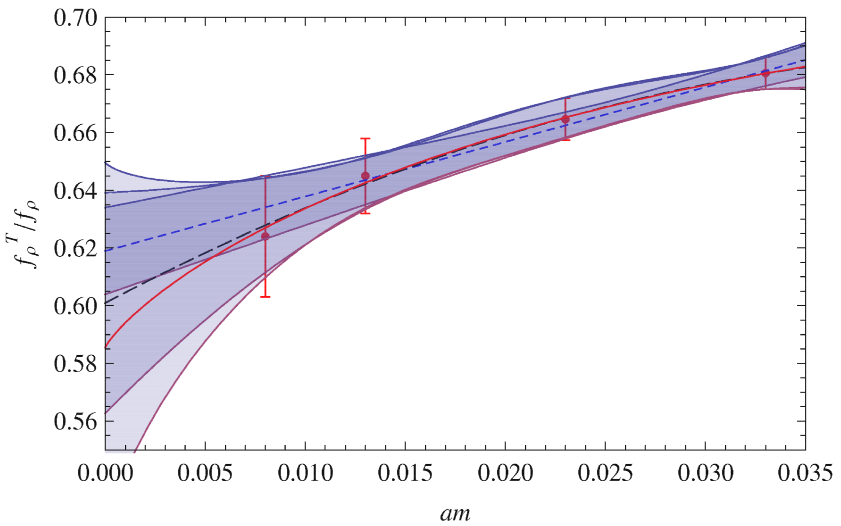}
\includegraphics[width=9.0cm]{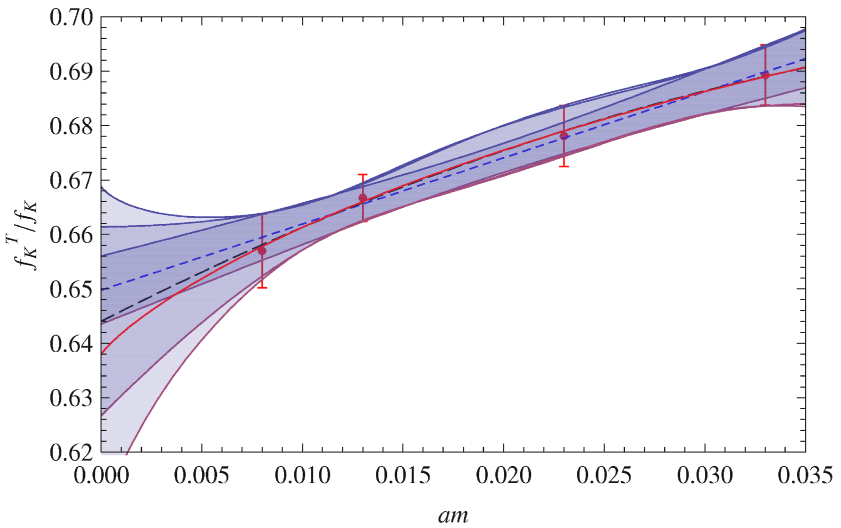}
\includegraphics[width=9.3cm]{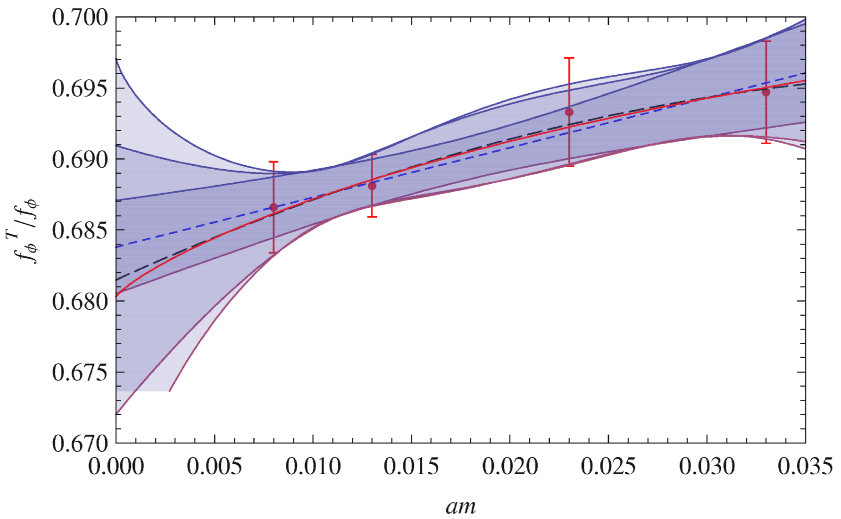}
\end{center}
\caption{Chiral extrapolations for the quadratic (blue dotted line), quartic (dashed line) and logarithmic (solid red line) fits with its corresponding error bars. Data taken from Ref.~\cite{Allton:2008pn}. In order to make the comparison with the results of Ref.~\cite{Allton:2008pn} easier, the plots show the quark mass (instead of the pion mass) dependence of $f_V^{\perp}/f_V$.   \label{fig:2} }
\end{figure}

To the best of our knowledge, the only unquenched lattice results available at present are the ones of Ref.~\cite{Allton:2008pn, Jansen:2009hr}. We will concentrate here on the values reported in Ref.~\cite{Allton:2008pn}. There one can find the values
for the ratio of tensor over vector decay couplings at four different values of the light quark masses. Their results are shown in Fig.~\ref{fig:2}. In principle one could now fit those data points with Eqs.~(\ref{su2}) and (\ref{results3}). However, there are two main objections to performing such a fit\,: first, the data in Ref.~\cite{Allton:2008pn} is not in the continuum limit, and actually at a rather big lattice spacing, $a^{-1}=1.729(28)$ GeV, so the use of $\chi$PT is, strictly speaking, not justified; second, data is so scarce for each channel and so far away from the chiral limit that the statistical significance of the fit would be extremely poor.

\begin{table}[tbh]\label{tab:comp}
\renewcommand{\arraystretch}{1.3}
\caption{Results for $f_V^{\perp}/f_V$ from the various fits (data taken from Ref.~\cite{Allton:2008pn}). Note that in contrast to Ref.~\cite{Allton:2008pn}, we have named the fits according to their scaling with the pion mass.}
\begin{center}
\begin{tabular}{ccccc}
\hline 
 & ${\mathrm{Quadratic}}$ & ${\mathrm{Quartic}}$ & ${\mathrm{Cubic}}$ & ${\mathrm{Logarithmic}}$ \\
\hline \hline $f_{\rho}^{\perp}/f_{\rho}$ & $0.619(15)$ & $0.600(39)$ & $0.595(47)$ & $0.585(65)$ \\ 
\hline $\chi^2/{\mathrm{dof}}$ & $0.17$ & $0.09$ & $0.07$ & $0.06$ \\
\hline\hline
$f_K^{\perp}/f_K$ & $0.6496(63)$ & $0.644(18)$ & $0.642(22)$ & $0.638(32)$ \\ 
\hline $\chi^2/{\mathrm{dof}}$ & $0.11$ & $0.09$ & $0.08$ & $0.07$ \\
\hline\hline
$f_{\phi}^{\perp}/f_{\phi}$ & $0.6837(33)$ & $0.6814(96)$ & $0.681(12)$ & $0.680(17)$ \\ 
\hline $\chi^2/{\mathrm{dof}}$ & $0.10$ & $0.12$ & $0.13$ & $0.15$ \\
\hline\hline
\end{tabular}
\end{center}
\end{table}

Until more data is available, one interesting exercise one can perform is to make a blind analysis of the lattice data in order to test its scaling with the pion mass. In particular, we want to test if lattice data comply with our results in Eq.~(\ref{results3}), favouring the absence of chiral logarithms in the $K^*$ and $\phi$ channels. We have performed two different fits on the lattice data, one with a quadratic and cubic dependence on the pion mass (the cubic fit), and one with a quadratic and chiral logarithmic dependence (the logarithmic fit), and compared them with the quadratic and quartic fits of Ref.~\cite{Allton:2008pn}.\footnote{Note that in Ref.~\cite{Allton:2008pn} fits are named according to its scaling with the light quark masses, whereas in this work the terminology for the fits refers to its pion mass scaling.} The results for the various fits are collected in Table~2, where for comparison we have also included the value of the chirally-extrapolated ratios $f_V^{\perp}/f_V$. Two comments are in order at this point\,: (i) in performing this exercise we are implicitly assuming that the qualitative behaviour of $f_V^{\perp}/f_V$ as a function of the quark masses is rather insensitive to the lattice spacing, such that Eq.~(\ref{results3}) can be used. This assumption is however not so implausible: similar exercises in the light pseudoscalar sector seem to indicate that, even away from the continuum limit, chiral logarithms are correctly reproduced; (ii) we should emphasize that the chirally-extrapolated values for the ratios listed in Table~2 should not be taken as actual values for $f_V^{\perp}/f_V$. Rather, they should be considered only as an indication of how the different fits would affect the predictions for $f_V^{\perp}/f_V$\,: as already noted above, the results reported in Ref.~\cite{Allton:2008pn} are not in the continuum limit.

Based on the $\chi^2$/dof for the different fits, the results of our analysis show that $\phi$ data seem to prefer the quadratic behaviour over the logarithmic one. In contrast, for the $\rho$ data the best fit is the logarithmic one, although only in a marginal way. Data seems to describe the characteristic bending of a logarithmic behaviour, but the statistical errors are still too big to be conclusive. Therefore, in the absence of better statistics no firm conclusions can be drawn at this point. The situation for the $K^*$ meson is less clear, but preliminary data from ETMC at three different lattice spacings seem to strongly prefer the quadratic fit over the logarithmic one.\footnote{Petros Dimopoulos, private communication.} Therefore, lattice data, at least at a qualitative level, seems to agree remarkably well with the predictions coming from chiral symmetry. 
   
\section*{Acknowledgments}
We would like to thank Jonathan Flynn and Andreas Juettner for correspondence and Petros Dimopoulos and Tassos Vladikas for useful discussions at the different stages of this work. O.~C. is also grateful to the Max-Planck-Institut f\"ur Physik (Werner-Heisenberg-Institut) for hospitality. This work is supported by the EU under contract MTRN-CT-2006-035482 Flavianet.


\begin{thebibliography}{99}
\bibitem{Ball:2006eu}P.~Ball, G.~W.~Jones and R.~Zwicky,      Phys.\ Rev.\  D {\bf 75}, 054004 (2007).

\bibitem{Becirevic:2003pn}D.~Becirevic, V.~Lubicz, F.~Mescia and C.~Tarantino, JHEP {\bf 0305}, 007 (2003).

\bibitem{Braun:2003jg}V.~M.~Braun, T.~Burch, C.~Gattringer, M.~Gockeler, G.~Lacagnina, S.~Schaefer and A.~Schafer,  Phys.\ Rev.\  D {\bf 68}, 054501 (2003).

\bibitem{Gockeler:2005mh}M.~Gockeler {\it et al.},   PoS {\bf LAT2005}, 063 (2006)   [arXiv:hep-lat/0509196].

\bibitem{Donnellan:2007xr}M.~A.~Donnellan {\it et al.},  PoS {\bf LAT2007}, 369 (2007)   [arXiv:0710.0869 [hep-lat]].

\bibitem{Dimopoulos:2008ee}P.~Dimopoulos, C.~McNeile, C.~Michael, S.~Simula and C.~Urbach  [ETM Collaboration], arXiv:0810.1220 [hep-lat].

\bibitem{Dimopoulos:2008hb}P.~Dimopoulos {\it et al.},   PoS {\bf LATTICE2008}, 271 (2008)   [arXiv:0810.2443 [hep-lat]].

\bibitem{Allton:2008pn}C.~Allton {\it et al.}  [RBC-UKQCD Collaboration], arXiv:0804.0473 [hep-lat].

\bibitem{Jansen:2009hr}K.~Jansen, C.~McNeile, C.~Michael, C.~Urbach and f.~t.~E.~Collaboration, arXiv:0906.4720 [hep-lat].

\bibitem{Jenkins:1995vb}E.~E.~Jenkins, A.~V.~Manohar and M.~B.~Wise, Phys.\ Rev.\ Lett.\  {\bf 75}, 2272 (1995).

\bibitem{Mateu:2007tr}V.~Mateu and J.~Portoles,   Eur.\ Phys.\ J.\  C {\bf 52} (2007) 325.

\bibitem{Jamin:2008rm}M.~Jamin and V.~Mateu, JHEP {\bf 0804} (2008) 040.

\bibitem{Ball:2002ps}P.~Ball, V.~M.~Braun and N.~Kivel, Nucl.\ Phys.\  B {\bf 649} (2003) 263.

\bibitem{Cata:2008zc}O.~Cata and V.~Mateu,  Phys.\ Rev.\  D {\bf 77}, 116009 (2008).

\bibitem{Bijnens:1997ni}J.~Bijnens, P.~Gosdzinsky and P.~Talavera, Nucl.\ Phys.\  B {\bf 501}, 495 (1997).

\bibitem{Bijnens:1998di}J.~Bijnens, P.~Gosdzinsky and P.~Talavera,   Phys.\ Lett.\  B {\bf 429}, 111 (1998).

\bibitem{Cata:2007ns}O.~Cata and V.~Mateu,   JHEP {\bf 0709} (2007) 078.
\end{thebibliography}
\end{document}